\documentclass{revtex4-1}
\usepackage[utf8]{inputenc}
\usepackage[T1]{fontenc}
\usepackage{setspace}
\usepackage{graphicx}
\usepackage{amsmath}
\usepackage{caption}

\begin{document}
\title{Entropically Dominant State of Proteins}

\author{Wenzhao Li}
\affiliation{College of Life Science, Jilin University 2699 Qianjin Street Changchun China 130012}
\author{Kai Wang}
\affiliation{College of Life Science, Jilin University 2699 Qianjin Street Changchun China 130012}
\author{Suyan Tian}
\affiliation{Division of Clinical Epidemiology, First Hospital of The Jilin University, 71 Xinmin Street, Changchun, Jilin, China. 130021}
\email{windytian@hotmail.com}
\author{Pu Tian}
\affiliation{College of Life Science}
\affiliation{Key Laboratory of Molecular Enzymology and Engineering of the Ministry of Education \\
Jilin University 2699 Qianjin Street Changchun China 130012}
\email{tianpu@jlu.edu.cn}

\begin{abstract}
Configurational entropy is an important factor in the free energy change of many macromolecular recognition and binding processes, and has been intensively studied. Despite great progresses that have been made, the global sampling remains to be a grand challenge in computational analysis of relevant processes. Here we propose and demonstrate an entropy estimation method that is based on physical partition of configurational space and can be readily combined with currently available methodologies. Tests with two globular proteins suggest that for flexible macromolecules with large and complex configurational space, accurate configurational entropy estimation may be achieved simply by considering the entropically most important subspace. This conclusion effectively converts an exhaustive sampling problem into a local sampling one, and defines entropically dominant state for proteins and other complex macromolecules. The conceptional breakthrough is likely to positively impact future theoretical analysis, computational algorithm development and experimental design of diverse chemical and biological molecular systems. 
\end{abstract}

\maketitle

\section{Introduction}
Configurational entropy (CE) plays an important role in many molecular recognition processes such as protein-protein interactions and protein-ligand binding\cite{Gilson2007,Demers2009,Marlow2010,Itoh2010,Tzeng2012}. Recent findings on the significant role of various level of disorder in protein functions are rapidly changing the traditional structure-function paradigm\cite{Tompa2012}, and are suggesting increasing urgency for quantitatively investigating entropic contributions in relevant free energy analysis. NMR spectroscopy, when combined with calorimetry, is a very powerful tool in elucidating the role of entropy\cite{Demers2009,Tzeng2012}. However, experimental characterization of CE is costly and many macromolecular systems are not amenable to NMR. Computational approach is a potentially economical and rapid alternative in evaluating CE\cite{Gilson2007}. In this regard, significant progress has been made since the seminal work by Karplus and Kushick\cite{Karplus1981} on estimation of CE with quasiharmonic approximation (QHA), which gives an upper bound of the true CE. The error of entropy estimated by QHA has three origins. Firstly, when Cartesian coordinates were used for calculation of the covariance matrix, overall rotation and translation may exaggerate the extent of molecular motion and consequently results in a larger entropy. This problem can be alleviated by using internal coordinates\cite{Harpole2011}. Secondly, the actual anharmonicity of each mode (i.e. non-Gaussian probability distributions along each mode) always exists. Thirdly, the matrix diagonalization process removes linear correlations and other correlations still exist between different modes. Many improvements\cite{Andricioaei2001,Baron2009,Harpole2011,Su2011,Nguyen2012} of QHA have been carried out to alleviate these problems in calculation of macromolecular CE. Mutual information expansion (MIE) based methodologies\cite{Chang2003,Chang2004,Chang2005,Hensen2010,Numata2012} were developed to account for non-linear correlations that are not accounted for in simple QHA type of methods. Another recent addition is the maximum spanning information tree (MIST) approximation\cite{King2012} that was shown to compare favourably with MIE approaches.

Partition of configurational space has been utilized by separating the full CE into conformational entropy and vibrational entropy of important conformers as shown below\cite{Karplus1987, Chang2005}:
\begin{equation}
S_{config} = \langle{S_i}\rangle - k_B\sum\limits_{i}P_iln(P_i)
\label{eq:1}
\end{equation}
Here $i$ represent individual conformational ensembles (correspond to configurational subspaces), $P_i$ and $S_i$ are the probability and vibrational entropy of the subspace (conformer) $i$, respectively. This idea was initially proposed for analysis of entropy change of proteins upon folding, and later on utilized in analysis of CE in protein-ligand interactions\cite{Chang2007, Chiba2012} and protein-protein associations\cite{Su2011}. A reported alternative is to use entropy invariant transformations to separate the configurational space into minimally coupled subspaces\cite{Hensen2010}. Implicit in both formulation, and deeply engrained in our minds, is that global sampling is essential to achieve accurate CE estimation. In this study, we provide an alternative formulation that relates total CE to CE of configurational subspaces and demonstrate its application in estimating CE of proteins from molecular dynamics (MD) trajectories. In our formulation, local CE of the most important (entropically) configurational subspace is proved to be a sufficiently accurate substitute of the total CE. This conclusion effectively transforms the widely accepted global sampling problem into a local sampling one, and suggests the existence of the entropically dominant state (conformer)[EDS] for proteins.

\section{Results}
\subsection{Theory}

In statistical mechanics, entropy is a logarithmic measure of the density of states:
\begin{equation}
\label{eq:2}
S = -k_{B} \sum_{j} {P_jln(P_j)}
\end{equation}
Where $k_B$ is the Boltzmann constant, the summation is over all possible \emph{microstate} and $P_j$ is the probability of the microstate $j$(note here it is the microstate, not conformations mentioned in eq. \ref{eq:1}). It is well established in informational entropy theory that $-\sum{P_iln{P_i}}$ is maximized for a uniform distribution\cite{Entropy}. Therefore, for a general ensemble,
\begin{equation}
\label{eq:3}
S \leq S_{max} = k_Bln\Omega
\end{equation} 
with $\Omega$ being the total number of microstates. The equality holds for microcanonial ensemble (a system with fixed number of particles, volume and energy), and that is the fundamental postulate of statistical mechanics. Consequently, Eq. \ref{eq:3} can be readily used as an upper bound for estimating CE. If we partition the configurational space into \emph{non-overlapping} and \emph{complete} $n$ subspaces $\Omega = \Omega_1 + \Omega_2 + \cdots + \Omega_n$, then we have:
,
\begin{equation}
\label{eq:4}
S_{max} = k_Bln(\Omega_1 + \Omega_2 + \cdots + \Omega_n) \\
\end{equation}
and
\begin{equation}
\label{eq:5}
S_{max1} = k_Bln\Omega_1, S_{max2} = k_Bln\Omega_2, \cdots, S_{maxn} = k_Bln\Omega_n
\end{equation}

After a few algebraic steps, calculation of the total CE can be transformed into estimation of entropy in configurational subspaces as shown below:

\begin{equation}
\label{eq:6}
S_{max} = S_{max1} + k_B ln(1 + e^{\frac{S_{max2}-S_{max1}}{k_B}} + \cdots))
\end{equation}

In any ensembles that are not microcanonical, distribution of microstates is not uniform. However, for a molecular system in equilibrium under given thermodynamic conditions, both the total CE and the CE of a given subspace is a fixed quantity. The entropy $S_t$ of the subspace $t$ can be calculated from the true distribution $S_t = -\sum_{j=1}^{j=\Omega_t}P_jlnP_j$. Alternatively, it can be regarded as the entropy of an equivalent subspace with uniformly distributed microstates, with the number of microstates $\Omega_{t'} < \Omega_t$ so that $S_t = k_B ln\Omega_t'$. We will consequently have a tighter bound for the total CE as:
\begin{equation}
\label{eq:7}
S_{max'} = S_1 + k_B ln(1 + e^{\frac{S_2-S_1}{k_B}} + \cdots))
\end{equation}   

 Where $S_i(i=1\cdots n)$ can be calculated with any available methods for estimating configurational entropy. A major difficulty in estimating configurational entropy of biomolecules (especially proteins) is the sampling problem. In Eq.\ref{eq:3}, each individual microstate contribute to the total entropy by adding $1$ to the $\Omega$. This requires complete sampling for \emph{brute force} estimation of configurational entropy. However, we note that $ln(N)$ is a very slowly increasing function of $N$, especially when $N$ is large, and this property is not used to our advantage in \emph{brute force} entropy estimation. Our reformulation in Eq.\ref{eq:5} utilized this fact. If we order the subspace entropy terms such that $S_1 \geq S_2 \geq \cdots \geq S_n $, it is immediately obvious that under the following conditions: i) $n$ is not a very large number, ii) $S_1$ is much larger than $k_B$ and iii) $S_1 - S_2$ is larger than $k_B$, $S_1$ would be a very good approximation of $S$. In practice, these conditions can almost always be satisfied for flexible macromolecules such as proteins and contributions from smaller terms ($S_2, S_3, \cdots, S_n$) become negligible. This formulation implies that if the most important subspace was sampled adequately, an accurate estimation of configurational entropy is achievable without global sampling. The most important concept from this formula is existence of \emph{EDS} for proteins, a stark challenge to our traditional association of entropy calculation with global sampling. One immediately arising question is that Eqs.\ref{eq:4} and \ref{eq:5} do not specify how to partition configurational space of a given molecular system, one may concern that different ways of partition lead to different CE estimation, thus prevent practical application of this formulation.  Additionally, complex biomolecules (e.g. proteins) have hierarchical free energy landscape (FEL), therefore, partition according to free energy wells need to have specific time scales, which may also be a potential source of complication. However, as Eqs.\ref{eq:4}, \ref{eq:5} and \ref{eq:6} are recursively true, as along as the number of subspaces is under control (significantly smaller than quantum mechanically allowed number of microstates in the EDS), this formulation will be effective regardless of specific ways to partition the configurational space.

\subsection{Tests with globular proteins}
\begin{figure}[p]
\centering
\includegraphics[angle=-90,width=12cm]{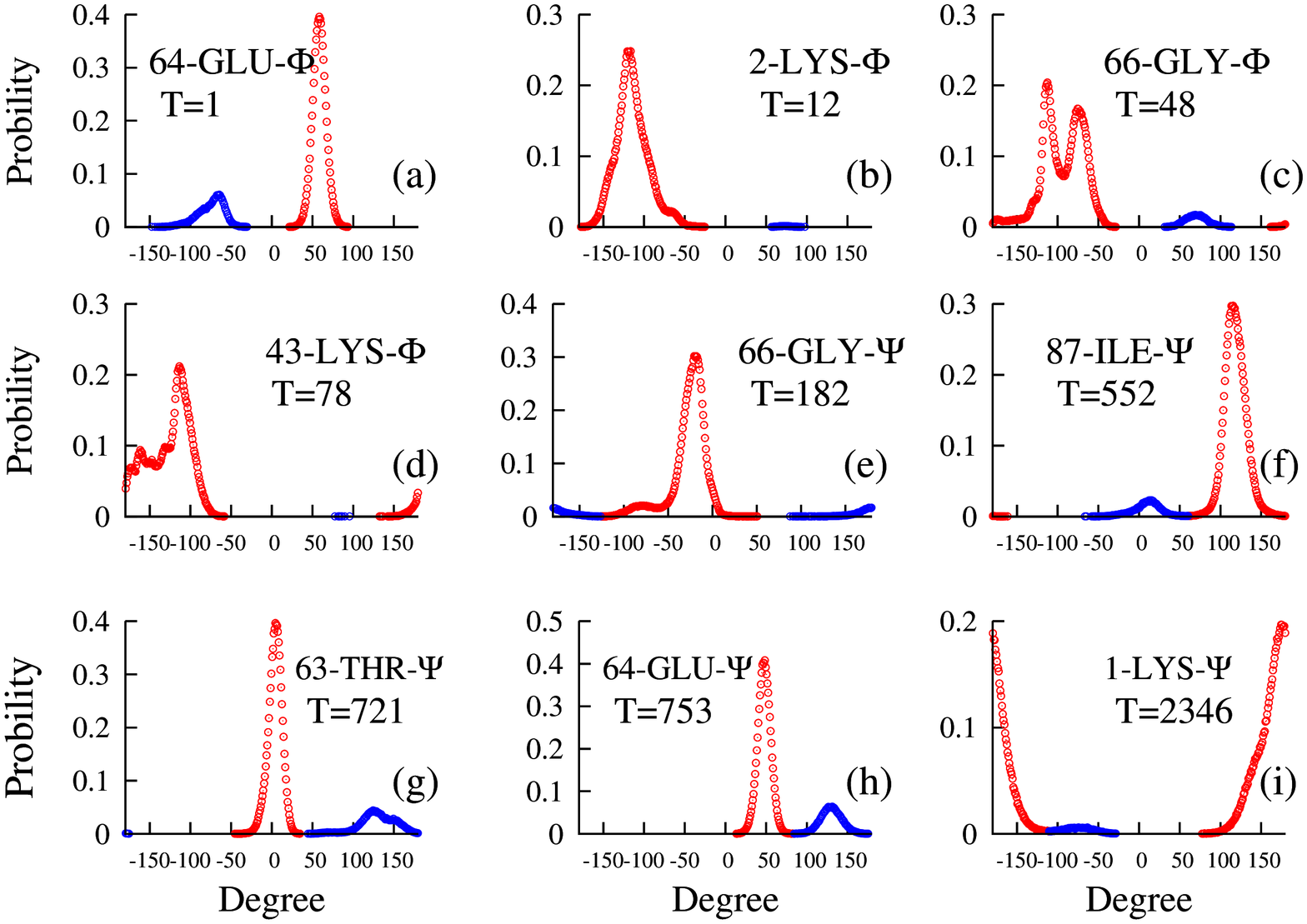}
\caption{Distributions of indicated backbone dihedral angles (64-GLU-$\phi$ indicate the $\phi$ angle associated with the 64th residue GLU.) that are utilized in partition of configurational space for barstar. Each color (red or blue) represent for a partition in the shown dihedral angle. $T$ is the number of transitions observed between different partitions.}
\label{Fig:1}
\end{figure}

\begin{figure}[p]
\centering
    \begin{tabular}{cc}
      \resizebox{60mm}{!}{\includegraphics[angle=-90]{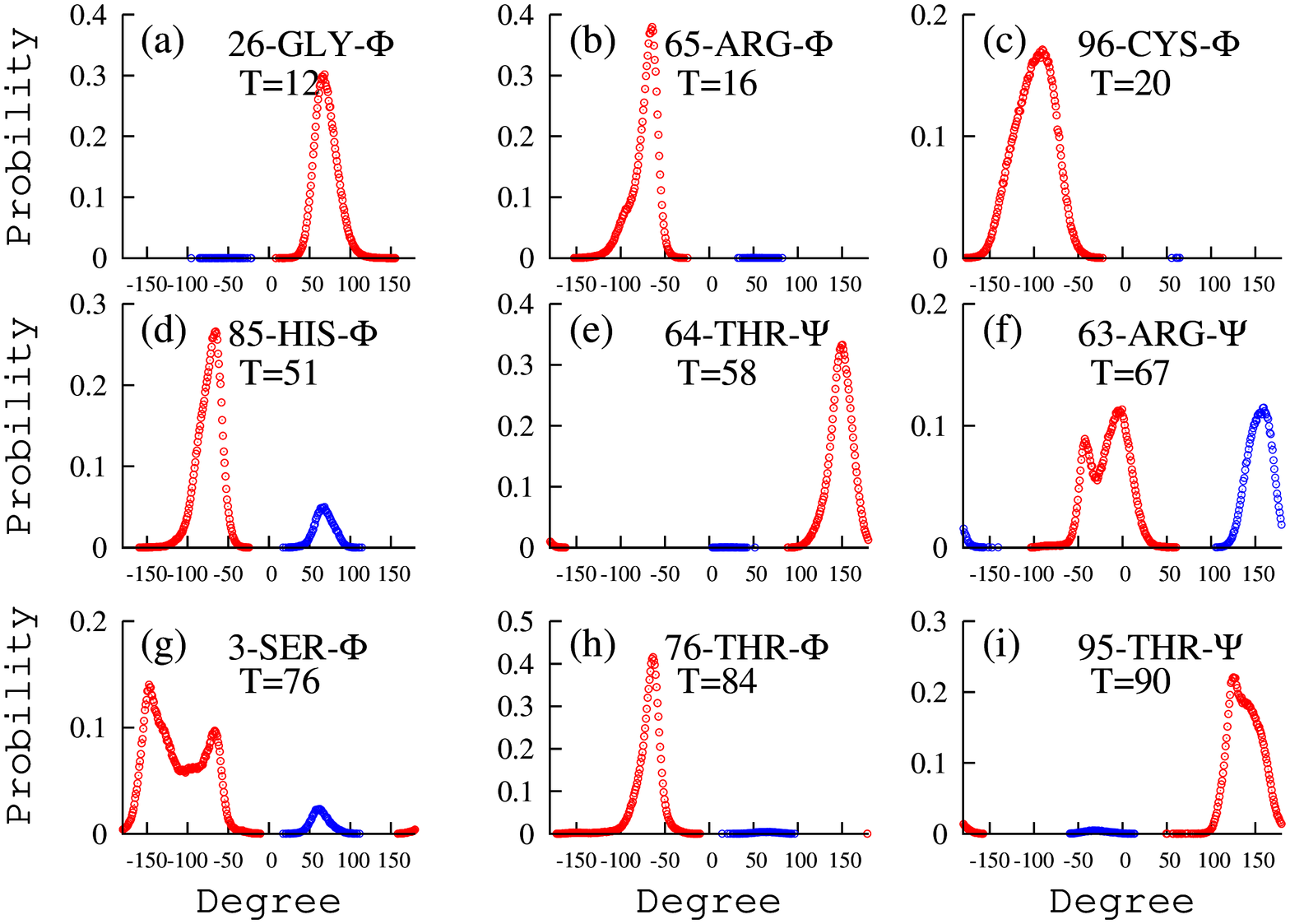}} &
      \resizebox{60mm}{!}{\includegraphics[angle=-90]{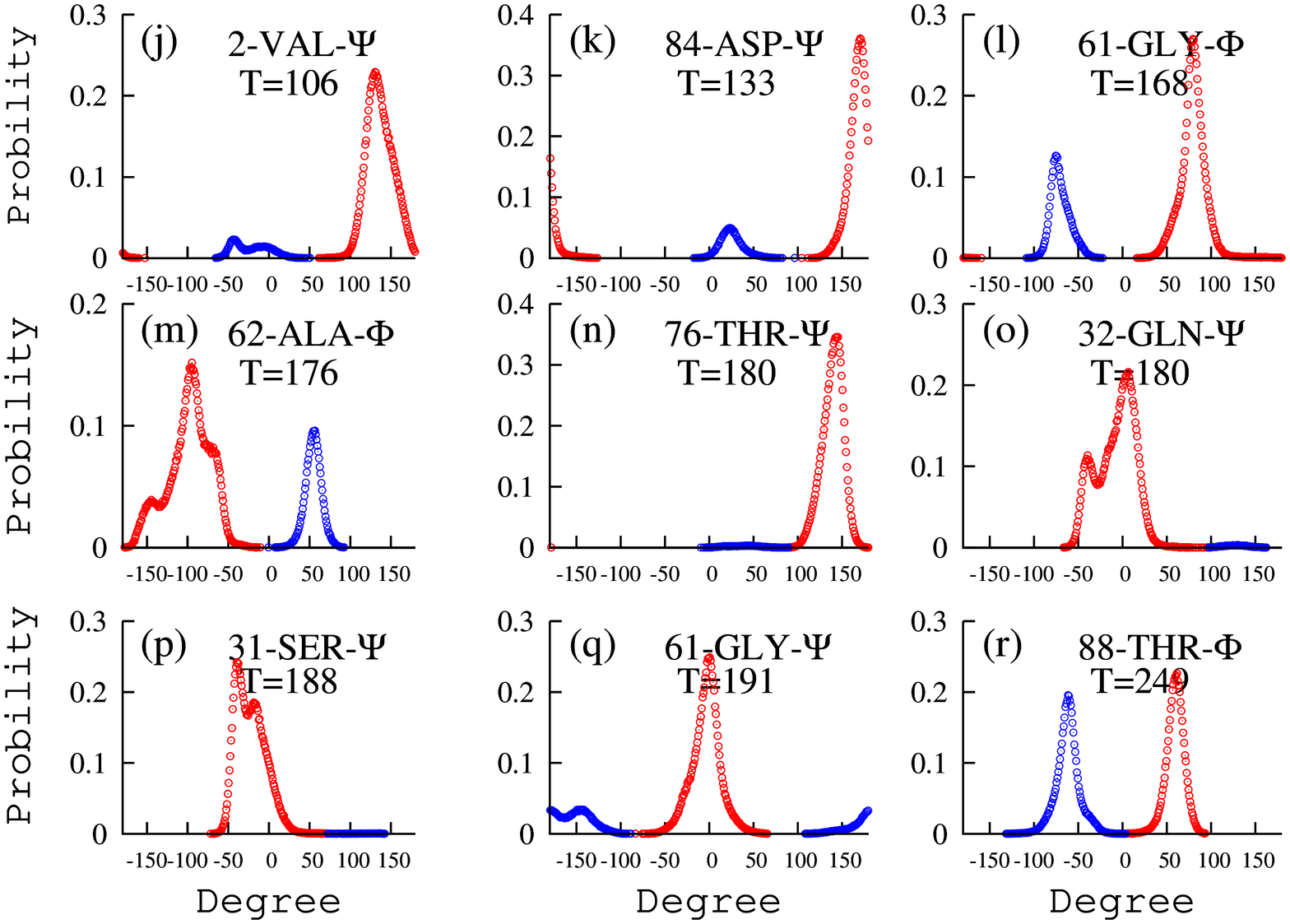}} \\
      \resizebox{60mm}{!}{\includegraphics[angle=-90]{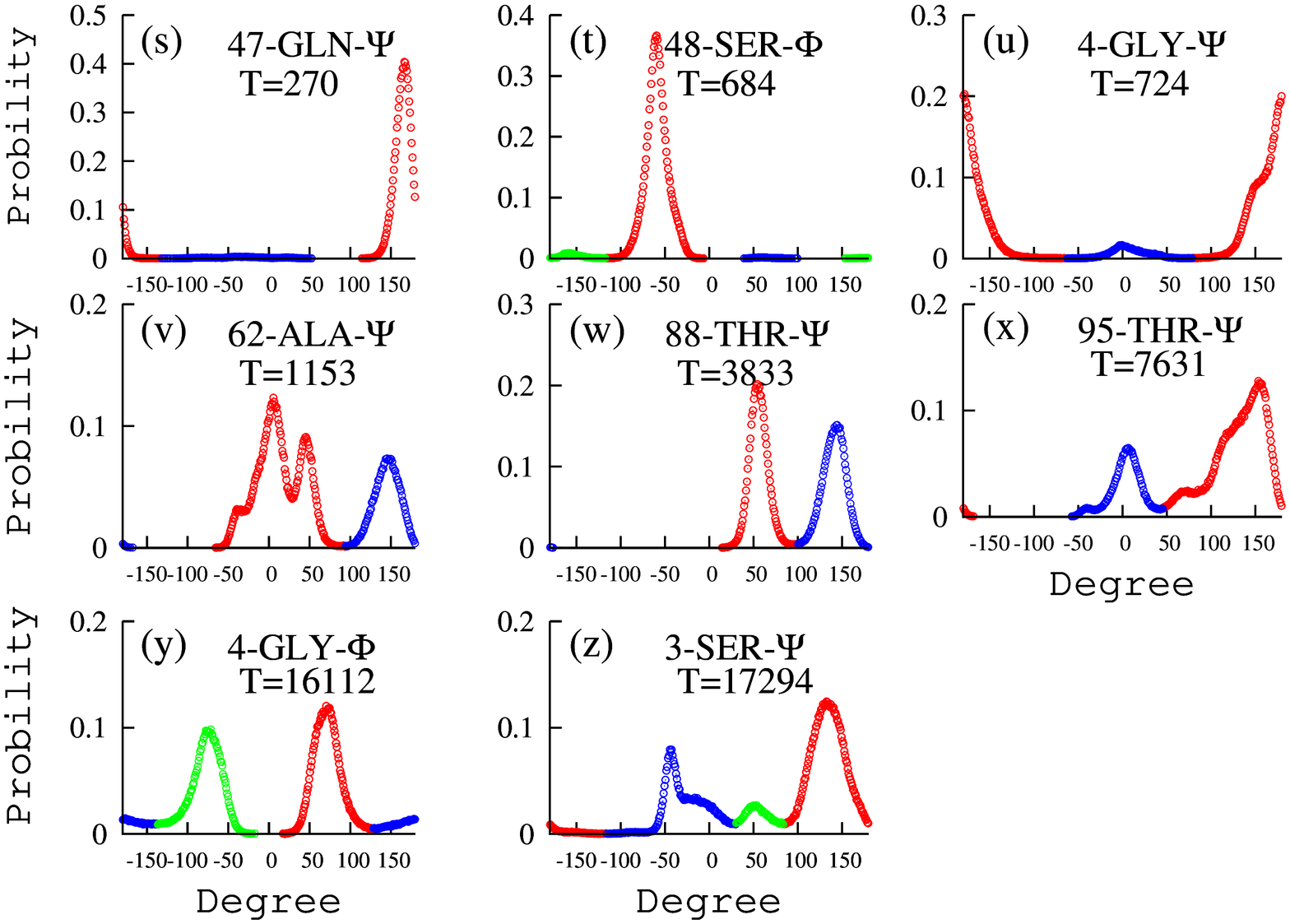}} &
    \end{tabular}
\caption{Distributions of indicated backbone dihedral angles that are utilized in partition of configurational space for the ribonuclease. Each color (red, blue or green) represent for a partition in the shown dihedral angle. $T$ is the number of transitions observed between different partitions.}
\label{Fig:2}
\end{figure}

To demonstrate the utility of configurational space partition in calculation of protein configurational entropy, we first generated 500-$ns$ long MD simulation trajectories for two globular proteins. One is a small ribonuclease protein from \emph{Streptomyces aureofaciens} and the other is its inhibitor barstar. While there are many possible ways for partition of configurational space, here as an example, we utilized distribution of backbone dihedral angles to carry out the partition. Entropy in each subspace is estimated with the widely used quasiharmonic calculations. Specifically, distributions of all backbone dihedral angles are plotted and those with multiple peak distributions were considered to be used for configurational space partition (see Fig. \ref{Fig:1}), and partitions were carried out in different hierarchies according to the time scale of relevant dihedral angle transitions between minima. 
After configurational space partition, the overall translation and rotation of snapshots within each subspace were removed through a fitting procedure (as widely used in RMSD calculations). Subsequently, covariance matrices $C_{i,j}=\langle(x_i - \langle(x_i)\rangle)(x_j - \langle x_j \rangle)\rangle $ were generated for subspaces $\Omega_1, \Omega_2, \Omega_3, \cdots, \Omega_n$ and entropy terms $S_1, S_2, S_3, \cdots, S_n$ were calculated according to the Schlitter formula\cite{Schlitter1993}. The calculated entropy and the number of snapshots in each subspaces were plotted for both proteins as shown in Fig. \ref{Fig:2}, The decrease of CE as a function of partition level is due to the improvement of the quasiharmonic approximation, which gives an upper bound for configurational entropy. As the configurational space is partitioned into finer subspaces, quasiharmonic approximation becomes better and thus gives tighter upper bounds ( smaller configurational entropy values ). It is apparent that for both proteins at each level of partition, the largest term dominate and contributions from other subspaces are negligible, confirming our initial speculation. Theoretically, only are the number of subspaces ($n$) comparable to the number of quantum mechanically allowed microstates in EDS, contributions from smaller terms ($S_2, S_3, \cdots, S_n$) become significant. For two protein investigated here, on the shortest time scale of partition 
more than one thousand subspaces were generated for the ribonuclease and only the largest term dominate. In practice, one rarely need to divide the configurational space into thousands of subspaces as that would result in many states that are not functionally meaningful. Therefore, this approximation will always be sufficiently accurate for practical applications. 

It is argued in the \emph{theory} section that Eqs. \ref{eq:4}, \ref{eq:5}, \ref{eq:6} and \ref{eq:7} are recursively true. A direct corollary is that when the configurational partitions were carried out in multiple hierarchies, the EDS at a given level must be a subspace from the EDS of its parent hierarchy. This is indeed the case for the conformational space of the two globular proteins that were explored by MD simulations. The EDS on the two lower hierarchies in Fig. \ref{Fig:2} are subspaces of the EDS from the corresponding parent hierarchy.  

\subsection{Comparison with other CE estimation methods}
\begin{figure}[p]
\centering
\includegraphics[angle=-90,width=12cm]{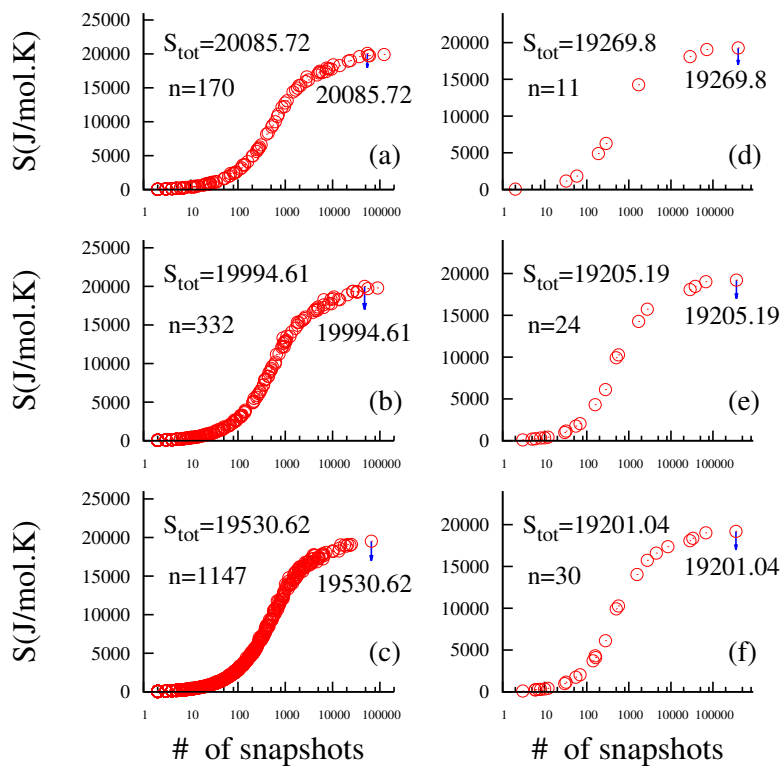}
\caption{The configurational entropy of subspaces as a functional of the number of snapshots within corresponding subspaces. $S_{tot}$ is the totoal CE, $S_1$ is pointed out with a blue arrow. $n$ is the number of subspaces for a given partition scheme. a), b) and c) correspond to the three partition hierarchies for the ribonuclease, dihedral angles used are Fig.\ref{Fig:2}a - Fig.\ref{Fig:2}u, Fig.\ref{Fig:2}a - Fig.\ref{Fig:2}x; and Fig.\ref{Fig:2}a - Fig.\ref{Fig:2}z respectively; d), e) and f) correspond to the three partition hierarchies for barstar, dihedral angles used are Fig.\ref{Fig:1}a - Fig.\ref{Fig:1}e, Fig.\ref{Fig:1}a - Fig.\ref{Fig:1}h; and Fig.\ref{Fig:1}a - Fig.\ref{Fig:1}j respectively.}
\label{Fig:3}
\end{figure}

Besides defining EDS for flexible macromolecules, our formulation has a few other advantages. When compared with theorerically rigorous mutual information expansion based methods, our formulation do not need to calculate high order correlation on a global scale. The partition of configurational entropy into vibrational and conformational components is based on the assumption that very few states are shared by different conformations (Fig. \ref{Fig:3}a), however, the widely accepted hierarchical FEL theory suggests that for a native protein, free energy barriers among the first tier of conformations are generally lower than folding/unfolding barriers, and this is recursively true when finer divisions (on shorter time scales) are considered. Such awkward situations were addressed by limiting relevant torsional DOF within $30^o$ of the minima\cite{Chang2007}(as shown in Fig. \ref{Fig:3}b), resulting in omission of many states close to transitional regions. In our formulation, partition is \emph{complete} and \emph{non-overlapping} (Fig. \ref{Fig:3}c), therefore possible contributions from transitional microstates are correctly counted. The conformational term is usually negligible when compared with the vibrational term in Eq. \ref{eq:1}, which is sometimes simplified to be the weighted average of vibrational entropies of individual conformation $S_{tot} = -\sum_i{P_i}ln{P_i}$\cite{Chiba2012}. This generates a paradox that the conformation with the largest vibrational entropy will have larger entropy than all the conformations combined. Our formulation is consistent in that the total CE corresponding to the whole configurational space is always slightly larger (by the negligible term in Eq. \ref{eq:7}) than the largest entropy ($S_1$ in Eq. \ref{eq:7}) of the comprising subspaces. Finally, in a regular simulation, extent of sampling is not uniform across all conformations, such practical sampling imbalance causes mixing of good (from very well sampled subspaces) and not so good (from insufficiently sampled subspaces) data. In contrast, our formulation focus on the EDS of a protein, usually located within the most well-sampled subspaces.

\section{Discussion}
It is worthwhile to note that the subspaces in eq. \ref{eq:3} correspond not necessarily to natural free energy wells. Any consistent partition of configurational space may be utilized. However, for the convenience of calculating CE in relevant subspaces, correspondence between configurational subspaces and free energy wells are helpful. One practical problem is that when a specific partition scheme is given, we do not know \emph{a priori} which subspaces are dominating for a given macromolecule or one of its major conformation. CE obtained from our two MD trajectories indicate that on a give time scale (FEL hierarchy), the number of snapshots in each subspace is roughly correlated with its configurational entropy. Therefore, one possible approach is to first using accelerated sampling methods\cite{Hu2012}(and references therein) to search for most important subconformations and subsequently perform enhanced local sampling within selected conformations that likely include the EDS. Specific experimental information (e.g. distance restraints from NMR, FRET or crosslink studies) that restricts configurational space can also be utilized to specify local sampling conditions.

In contrast to widely hold approximations that the number of available rotameric states dominates the entropy loss upon ligand binding and protein association, detailed entropy analysis for protein-ligand binding\cite{Chang2007} and protein-protein associations\cite{Chang2008} demonstrated that the narrowing of major free energy wells contributes the most. These observations obtained from complex analysis is an obvious conclusion of our formulation, which states that on a given free energy hierarchy, the EDS alone determines the CE of the molecule. 

It is important to note here that the EDS is not necessarily the state with the lowest enthalpy. The relationship between these two class of dominant substates may well be different for different macromolecules, and is an interesting topic to be studied in future investigations.

The global sampling problem has two aspects, one is to sample each significant free energy well as sufficiently as possible and the other is to sample as many free energy wells as possible. In a regular MD or Monte Carlo simulation, both the number of free energy wells visited on a given hierarchy and the extent of sampling in each visited well increase with increasingly longer trajectories. It is important to note that even if we ignored force field issues, sampling issue in CE calculation is not readily solved by running longer trajectories. The reason is that on the one hand, we do not know \emph{a priori} how long a trajectory is long enough, on the other hand, what we mostly interested in are CE differences between two different states of our target molecules (e.g. bound and free, or two different major conformations) and not the CE of one particular state(conformation). Within a given subspace, with increasing length of simulation trajectory, more and more of its configurational space will be visited and correspondingly larger CE will be obtained from the trajectory, therefore balanced sampling of relevant states (conformations) is essential. The conundrum is that if we knew \emph{a priori} the relative weight of two conformations in the partition function, we already knew the free energy difference and thus there is not much need to calculate the CE difference any more. If the two interested conformations repetitively occur in a single trajectory of simulation, then balanced sampling is generally considered to be satisfied if many (e.g. hundreds of) transitions were observed. However, for most of interested molecular events (e.g. ligand binding, formation of macromolecular complex, long time scale conformational changes), it is very difficult to achieve a balanced sampling with brute-force simulations. Free energy pathway analysis (e.g. string method\cite{Maragliano2006}) provides one possible solution with the limitation that one pathway (the minimum free energy pathway, MFEP) need to be the dominating transition path. It was proposed\cite{Numata2012} that to rapidly achieve convergence of entropy difference ($\Delta{S}$), equal number of snapshots need to be obtained for the two interested conformations. This strategy indeed accelerates convergence, but in most cases to a wrong value. The only special situation this method stays rigorously correct is that the two involved conformations have the same free energy. Therefore, efficient strategies for achieving balanced sampling of interested states within available computational resources is highly desired.

In conclusion, our formulation of configurational entropy naturally defines EDS for proteins and represents an important new direction in CE estimation, that is to search for the EDS rather than to sample as many conformations/states as possible, thus greatly reduce the complexity of CE calculation. As the formulation is true for any flexible macromolecules that have large and complex configurational space, the methodology is not limited to protein molecules, which we used for demonstration. We hope that this conceptional breakthrough find practical applications in free energy calculations, development of macromolecular theory and experimental design of diverse chemical and biological systems.

\section{Methods}
All MD simulations were performed with NAMD software package\cite{NAMD} (version 2.7) and CHARMM27 force fields. Ribonuclease(pdb code:1rgh) and barstar(pdb code:1bta) were solvated with TIP3P water molecules. The simulation systems were neutralized by adding 100mM $N_a^+$ and $Cl^-$ ions. Bond-lengths involving hydrogen atoms were constrained using the SHAKE algorithm, and the integration time step is set to 2 $fs$. A switch distance of $10$\AA\ and a cut off distance of $12$ \AA\ were used for non-bonded interactions. Particle Mesh Ewald was used to calculate the long-range electronic interactions. Both systems were minimized and then heated to $310 K$ with heavy atoms restrained, water molecules were equilibrated with 200-$ps$ runs in NVT ensemble. After that, restraints for protein heavy atoms were released, and the whole system was equilibrated in the NPT ensemble for another 4 $ns$. A frame with the volume value that is closest to the average volume obtained from NPT equilibration run was selected to start the next production runs which were performed in the NVT ensemble at $310 K$. A 500-$ns$ trajectory was generated for each protein. Coordinates were saved every $ps$ for analysis.

\section{Acknowledgement}
This research was partially funded by a start-up fund from Jilin University, by the National Science Foundation of China (Grant \#31270758) and by the Scientific Research Foundation for the Returned Overseas Chinese Scholars, State Education Ministry.

\clearpage
\bibliography{./entropy}
\bibliographystyle{achemso}

\clearpage

\clearpage

\clearpage

\clearpage
\begin{figure}[p]
\begin{center}
\includegraphics[width=12cm]{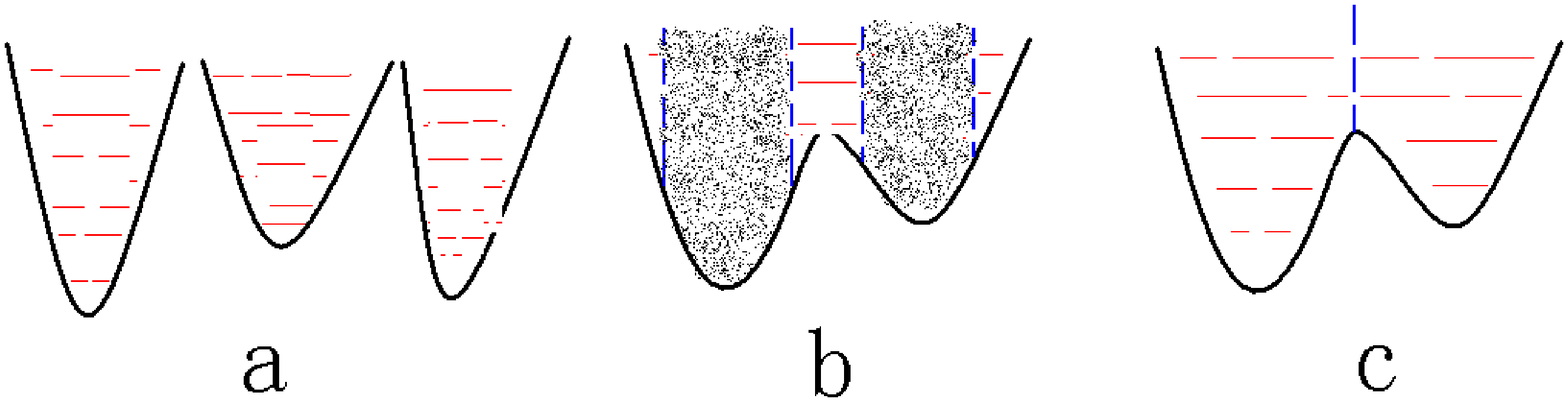}
\caption{Schematic representation of the configurational space partitions. a) Partition into individual conformations with transition region being neglectable; b) Transitional states between conformations have significant contribution to the configurational spaces, only shadowed area between blue lines were considered in traditional decomposition of CE into conformatioal entropy and vibrational entropy; c) a complete and non-overlapping partition with neither double counting nor omission, the vertical dashed blue line represent a possible border between two partitions.}
\label{Fig:4}
\end{center}
\end{figure}

\end{document}